\documentclass[aps,pra,twocolumn,groupedaddress,showpacs,amsmath,amssymb]{revtex4-1}

\usepackage{graphicx}
\usepackage{bm,color}

\usepackage{graphicx}
\usepackage{bm,color}

\newcommand{\be}{\begin{eqnarray}}
\newcommand{\ee}{\end{eqnarray}}

\renewcommand{\theequation}{\arabic{equation}}

\begin{document}

\title{Disorder-induced Chern insulator in Harper--Hofstadter--Hatsugai model}
\date{\today}

\author{Yoshihito Kuno} 

\affiliation{Department of Physics, Graduate School of Science, Kyoto University, Kyoto 606-8502, Japan}
\begin{abstract}
We study the effects of disorder on the topological Chern insulating phase in the Harper--Hofstadter--Hatsugai (HHH) model.
The model with half flux has a bulk band gap and thus exhibits a nontrivial topological phase. 
We consider two typical types of disorder: on-site random disorder and the Aubry--Andre type quasi-periodic potential. 
Using the coupling matrix method, we clarify the global topological phase diagram in terms of the Chern number.
The disorder modifies the gap closing behavior of the system. 
This modification induces the Chern insulating phase even in the trivial phase parameter regime of the system in the clean limit.
Moreover, we consider an interacting Rice--Mele model with disorder, which can be obtained by dimensional reduction of the HHH model. 
Moderately strong disorder leads to an increase in revival events of the Chern insulator at a specific parameter point. 
\end{abstract}

\maketitle
\section{Introduction}
How does disorder change the topological properties of a system?
This question has long been discussed and investigated.
Nontrivial topological phases such as the Chern insulator (CI) phase are generally said to be robust against weak disorder, 
because weak disorder cannot close the topological bulk band gap of the system. 
However, in systems with strong disorder, even if the symmetry of the topological model is not broken, 
the topological bulk band gap is closed. That is, the topological phase is maintained up to strong disorder. 
This topological robustness against disorder was first discussed by Thouless and colleagues \cite{Niu,Niu2}.
They suggested qualitatively that disorder and weak perturbation do not change the topological invariant number, 
i.e., the Chern number \cite{TKNN}, as long as they does not close the bulk band gap; thus, the topological state is robust against them. 
However, quantitative study of the robustness of various concrete topological models has not been completed. 

Counterintuitive behavior in some topological models was recently reported. 
Specifically, the topological phase may be induced by disorder. 
For example, in a three-dimensional topological insulator model, 
an interesting topological phenomenon induced by disorder, the topological Anderson insulator (TAI), 
has been theoretically suggested and numerically simulated \cite{Guo2,Groth,JLi,YYZhang,Jiang,ZQZhang}. 
In addition, in some typical two-dimensional (2D) topological models, 
both robustness and disorder-induced topological phenomena have been investigated 
\cite{Liu,Qin,YFZhang,Castro,Imura,Yamakage,Sriluckshmy,Prodan,Mondragon,Moscot}. 
Furthermore, very recently the Su--Schrieffer--Heeger model \cite{Shen,Asboth} was realized in an atomic quantum simulator \cite{Meier}. 
A momentum space lattice of cold atoms was designed to simulate the model, 
and the disorder-induced topological phase characterized by the Zak phase \cite{Zak} was observed.  
Motivated by this recent experimental success, in this study, we investigate the effects of disorder on the CI phase 
by considering the Harper--Hofstadter--Hatsugai (HHH) model \cite{Hatsugai,Zheng,Irgigler} as a concrete 2D topological model. 
The effects of disorder on this 2D topological model have not yet been discussed and estimated quantitatively. 
A complete understanding of the effects of disorder in the CI phase remains elusive.
If we apply dimensional reduction \cite{Kraus}, the HHH model is also important as an ancestral model of topological charge pumping \cite{Thouless}, 
i.e., the Rice--Mele (RM) model \cite{Rice}. Recently, there are some theoretical studies of effects of disorder in the RM model \cite{RWang,Ippoliti}. 
Accordingly, studying the effect of disorder on the CI phase in the HHH model 
is expected to afford important insight into the effect of disorder on topological charge pumping in the RM model.
\begin{figure}[b]
\centering
\includegraphics[width=6cm]{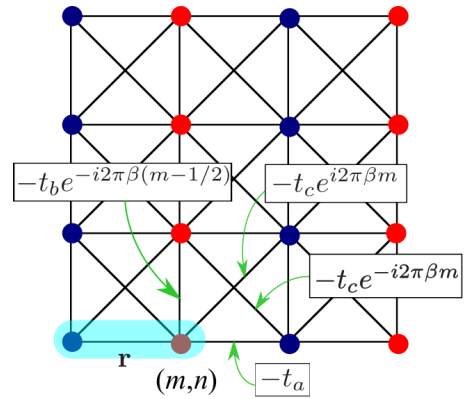}
\caption{Lattice structure of the HHH model. 
Red and blue circles represent sublattice sites for the half flux case ($\beta=1/2$).
The blue shaded cell is a unit cell at the position ${\bf r}$.}
\label{model}
\end{figure}
This paper is organized as follows: we first describe our target model in Sec.~\ref{SecII}. 
In Sec.~\ref{SecIII}, we briefly summarize a method of calculating the Chern number of a disordered system.
In Sec.~\ref{SecIV}, we clarify the global phase diagram and the robustness of the CI phase to disorder, 
and we study the enlargement of the CI phase.
Then we discuss the disorder-induced CI phase by calculating the density of states (DOS) in Sec.~\ref{SecV}. 
Further, in Sec.~\ref{SecVI}, we study the RM model associated with the HHH model. 
In particular, we study the effects of disorder on the CI phase in a bosonic interacting version of the RM model. 
Note that recent experimental realization of the RM model \cite{Lohse,Nakajima} motivates us to investigate the effects of disorder on the topological phase of the model.
Very recently, the bosonic RM model has been considered in some papers \cite{Qian,RLi,YKe,Hayward}. 
In this study, we consider phenomena outside of the conventional topological charge pumping trajectory. 
The observed tendency is the same as that in the non-interacting HHH model. 
We find that revival of the CI phase occurs with increasing strength of the random potential.
Finally, Sec.~\ref{SecVII} presents a discussion and conclusion.

\section{Model}\label{SecII}
The HHH model \cite{Hatsugai} is given by the following Hamiltonian:    
\begin{eqnarray}
H_{\rm 3H}&=&\sum_{m,n}\biggl[-t_{a}\hat{c}^{\dagger}_{m+1,n}\hat{c}_{m,n}
                 -t_{b}e^{-i2\pi\beta(m-1/2)}\hat{c}^{\dagger}_{m,n+1}\hat{c}_{m,n}\nonumber\\
                 &&-t_{c}e^{i2\pi\beta m}\hat{c}^{\dagger}_{m+1,n+1}\hat{c}_{m,n}
                 -t_{c}e^{-i2\pi\beta m}\hat{c}^{\dagger}_{m+1,n}\hat{c}_{m,n+1}\nonumber\\
                 &&-\Delta_{0}(-1)^{m}n_{m,n} +\mbox{h.c.}\biggr],
\label{HHH}
\end{eqnarray}
where $(m,n)$ is a lattice site on a 2D square lattice, $\beta$ is the flux per unit cell, and
$\hat{c}^{\dagger}_{m,n}$ ($\hat{c}_{m,n}$) is 
a fermionic creation (annihilation) operator at site $(m,n)$. Further, $n_{n,m}= c^{\dagger}_{n,m}c_{n,m}$ is a number operator, and $t_{a}$, $t_{b}$, and $t_{c}$ are the $x-$-direction, $y-$-direction, and diagonal hopping amplitudes, respectively. The hopping terms are shown in Fig.~\ref{model}. The
$\Delta_{0}$ term on the last line of Eq.~(\ref{HHH}) is a staggered potential that acts as a bulk-gap-controlling term. 
The model belongs to the AIII class in the topological classification scheme \cite{Schnyder,Kitaev}. For $\Delta_{0}=0$ and $t_{c} \neq 0$, 
the HHH model is known to have a nontrivial topological band structure corresponding to that of the Harper--Hofstadter model \cite{Hatsugai,Hofstadter}. 
While the Harper--Hofstadter model with $\beta=1/2$ has no bulk band gap, 
the HHH model with $\beta=1/2$ and $t_{c} \neq 0$ has a bulk band gap, and the band is topological. 
If the system is half-filled, the HHH model exhibits the CI phase.
Further, this bulk band gap is controlled by varying the value of $\Delta_{0}$. 
For $\Delta_{0}=2$, the bulk band gap is closed; then a topological phase transition occurs, and the system enters the trivial phase \cite{Zheng,Irgigler}.

We consider two diagonal types of disorder. The first is represented by the on-site random potential,
\begin{eqnarray}
V^{\rm I}_{d}=\sum_{(m,n)}\frac{{\mu}_{m,n}}{2}n_{m,n},
\label{dis1}
\end{eqnarray}
where the random potential has a uniformly random distribution with a strength $W_{0}$ that is defined as $\mu_{m,n}\in [-W_{0},W_{0}]$. 
The second is represented by an Aubry--Andre quasi-periodic potential defined on a 2D lattice,  
\begin{eqnarray}
V^{\rm II}_{d}=\sum_{(m,n)}\frac{V_{d}}{2}\biggl[\cos(2\pi \gamma m +\phi)+\cos(2\pi \gamma n +\phi)\biggr]n_{m,n},\nonumber\\
\label{dis2}
\end{eqnarray}
where $V_{d}$ is the potential amplitude, and $\gamma$ is set to the golden ratio, $(\sqrt{5}-1)/2$. 
We introduce a phase shift parameter $\phi$, which is convenient for suppressing the effect of boundary. 
We set $\beta=1/2$ and $t_{a}=t_{b}=t_{c}=1$, and we consider the half-filled case in the following study.

\section{Chern number from a real-space Hamiltonian}\label{SecIII}
We estimate the effects of disorder on the nontrivial topological phase in the HHH model.
Disorder breaks the system's translational symmetry, so bulk momentum representation cannot be employed. 
Thus, it is difficult to define and calculate the Chern number 
because its formula is calculated using the system wave function represented by momentum bases \cite{TKNN}. 
However, Niu and Thouless proposed another formula for the Chern number \cite{Niu,Niu2}, which is valid even in disordered systems. 
They introduced twisted boundary phase parameters under a periodic boundary condition, regarded these parameters as the quasi-momentum, 
and formulated the Chern number in terms of the parameters. 
By using this formulation, the Chern number can be defined even for systems that include disorder. 
However, to date it has generally been difficult to perform practical calculation of concrete lattice models in two spatial dimensions. 
On the basis of the work of Niu and Thouless, some recent papers \cite{Prodan,Bianco,Resta} have suggested 
the real-space version of the Chern number or the Berry phase; by using these proposed versions, some simple disordered topological models have been studied \cite{Prodan,Bianco,Irgigler}. 
As a simpler method, the coupling matrix method was recently proposed in \cite{YFZhang}. 
Although the method cannot extract local properties of the Chern number in a real space, called the local Chern marker \cite{Bianco}, 
it can extract the Niu and Thouless formula for the Chern number \cite{Niu,Niu2} correctly.

In this paper, we employ the coupling matrix method. 
Before proceeding, it is worth summarizing the coupling matrix method of practical numerical calculation in detail.

The coupling matrix method is based on the Niu and Thouless formula \cite{Niu,Niu2}.
The Chern number can be defined in terms of two twisted boundary phases, 
\begin{eqnarray}
C_{N}&=&\frac{1}{2\pi i}\int_{T^{2}}d{\bf \theta}\ \langle \nabla_{\bf \theta}\Psi({\bf \theta})|\times |\nabla_{\bf \theta}\Psi({\bf \theta})\rangle,
\label{CNphase}
\end{eqnarray}
where $|\Psi({\bf \theta})\rangle$ is the many-body wave function of the system, 
${\bf \theta}=(\theta_{x},\theta_{y})$ is the twisted boundary phase, and $\theta_{x}\; , \theta_{y}\in [0,2\pi)$. 
These phases are introduced using a twisted phase boundary condition for a single-particle wave function on a real-space lattice $\psi(r_x,r_y)$ as
 $\psi(r_{x}+L_{x},r_{y})=e^{i\theta_{x}}\psi(r_x,r_y)$ and $\psi(r_x,r_y+L_{y})=e^{i\theta_{y}}\psi(r_x,r_y)$, 
where ${\bf r}=(r_{x},r_{y})$ is a unit cell lattice site, as shown in Fig.~\ref{model}, and $L_{x(y)}$ is a piece length in the $x$ ($y$) direction scaled by the lattice spacing.
$T^{2}$ means a torus parameter space for ${\bf \theta}$. 
By applying the Stokes theorem to Eq.~(\ref{CNphase}), $C_{N}$ can be transformed into a line integral form.
Then we consider discretizing $T^{2}$ space (${\bf \theta}\to {\bf \theta}_{\ell}$) for numerical convenience later. 
The area integral of Eq.~(\ref{CNphase}) can be represented as 
\begin{eqnarray}
C_{N}=-\frac{1}{2\pi i}\prod_{{\bf \theta_{\ell}}\in \partial T^2}\langle \Psi({\bf \theta}_{\ell})|\Psi({\bf \theta}_{\ell+1})\rangle.
\label{CNphase2}
\end{eqnarray}
$\partial T^2$ is the boundary of the rectangular regime defined by ${\bf \theta}$ space. 
The right hand side is gauge-invariant, so we do not need to consider the gauge dependence of the wave functions obtained numerically \cite{Bernevig}.

Then, for non-interacting systems, the many-body wave function is given using single-particle states,    
\begin{eqnarray}
|\Psi({\bf \theta}_{\ell})\rangle = \frac{1}{\sqrt{N_{F}!}}
\begin{vmatrix}
|\psi^{1}_{{\bf \theta}_{\ell}}\rangle & \cdots &|\psi^{N_{F}}_{{\bf \theta}_{\ell}}\rangle\\
\vdots & \ddots  & \vdots \\
|\psi^{1}_{{\bf \theta}_{\ell}}\rangle & \cdots &|\psi^{N_{F}}_{{\bf \theta}_{\ell}}\rangle
\end{vmatrix},
\label{Cmatrix}
\end{eqnarray}
where $|\psi^{\alpha}_{{\bf \theta}_{\ell}}\rangle$ is the $\alpha$-th single-particle state with a twisted phase boundary condition ${\bf \theta}_{\ell}$, 
$\alpha=1,\cdots, N_{F}$. $N_{F}$ is the maximum number of occupied states, which is determined by the Fermi energy. 

By using the Slater determinant form of Eq.~(\ref{Cmatrix}), 
$\langle \Psi({\bf \theta}_{\ell})|\Psi({\bf \theta}_{\ell+1})\rangle$ in Eq.~(\ref{CNphase2}) is written as
\begin{eqnarray}
&&\langle \Psi({\bf \theta}_{\ell})|\Psi({\bf \theta}_{\ell+1})\rangle \equiv \det [A(\ell,\ell+1)], \label{detA}
\end{eqnarray}
where
\begin{eqnarray}
&&A(\ell,\ell+1)=
\begin{bmatrix}
\langle \psi^{1}_{{\bf \theta}_{\ell}}|\psi^{1}_{{\bf \theta}_{\ell+1}}\rangle & \cdots &\langle \psi^{1}_{{\bf \theta}_{\ell}}|\psi^{N_{F}}_{{\bf \theta}_{\ell+1}}\rangle\\
\vdots & \ddots  & \vdots \\
\langle \psi^{N_{F}}_{{\bf \theta }_{\ell}}|\psi^{1}_{{\bf \theta}_{\ell+1}}\rangle & \cdots & \langle \psi^{N_{F}}_{{\bf \theta}_{\ell}}|\psi^{N_{F}}_{{\bf \theta}_{\ell+1}}\rangle
\end{bmatrix}.
\label{Cmatrix2}
\end{eqnarray}
$A(\ell,\ell+1)$ is called the coupling matrix \cite{YFZhang}.

By substituting Eq.~(\ref{detA}) into Eq.~(\ref{CNphase2}), $C_{N}$ is given as
\begin{eqnarray}
C_{N}=-\frac{1}{2\pi i}\det \biggl[\prod_{{\bf \theta_{\ell}}\in \partial T^2}A(\ell,\ell+1)\biggr].
\label{CNphase3}
\end{eqnarray}
Consequently, to obtain $C_{N}$, 
the coupling matrix $A(\ell,\ell+1)$ needs to be calculated appropriately.

It is known that the elements of $A(\ell, \ell+1)$ can be obtained well only 
from single-particle spatial wave functions without twisted phase boundary conditions \cite{YFZhang}, i.e., ${\bf \theta}={\bf 0}$.
As shown in Ref. \cite{Niu,YFZhang}, when one introduces the nonzero twisted boundary phase, 
the real-space single-particle wave function is given in the following Fourier-transformed form:
\begin{eqnarray}
\langle {\bf r}|\psi^{\alpha}_{\bf \theta}\rangle =  \frac{1}{\sqrt{L_xL_y}}\sum_{{\bf k}\in BZ} e^{-i ({\bf k}+{{\bar{\bf\theta}}})\cdot {\bf r}}\langle {\bf k}|\psi^{\alpha}_{{\bf \theta}}\rangle,
\label{FTR10}
\end{eqnarray} 
where BZ represents the first Brillouin zone, and ${\bar{\bf\theta} }=(\theta_{x}/L_{x},\theta_{y}/L_{y})$.  
The above equation is surely satisfied under the twisted phase boundary conditions: $\psi^{\alpha}(r_x+L_{x},r_y)=e^{i\theta_{x}}\psi^{\alpha}(r_x,r_y)$, 
and $\psi^{\alpha}(r_x,r_y+L_{y})=e^{i\theta_{y}}\psi^{\alpha}(r_x,r_y)$. 
Here, by multiplying both sides of Eq.~(\ref{FTR10}) with $(\theta_{x},\theta_{y})={\bf 0}$ by $e^{-i{{\bar {\bf \theta}}_{\ell}}\cdot {\bf r}}$, we obtain 
\begin{eqnarray}
\langle {\bf r}| e^{-i{{\bar {\bf \theta}}_{\ell}}\cdot {\hat {\bf r}}}|\psi^{\alpha}_{{\bf \theta}={\bf 0}}\rangle &=& \frac{1}{\sqrt{L_xL_y}}\sum_{{\bf k}\in BZ} e^{-i ({\bf k}+{{\bar{\bf\theta}}_{\ell}})\cdot {\bf r}}\langle {\bf k}|\psi^{\alpha}_{{\bf \theta}}\rangle.\nonumber\\
\label{FTR5}
\end{eqnarray}
Note that the LHS of the above equation can be regarded as $\langle {\bf r}|\psi^{\alpha}_{{\bf \theta}_{\ell}}\rangle $ \cite{YFZhang}. 
Accordingly, we obtain the useful relation 
\begin{eqnarray}
e^{-i{{\bar {\bf \theta}}_{\ell}}\cdot {\hat {\bf r}}}|\psi^{\alpha}_{{\bf \theta}={\bf 0}}\rangle=|\psi^{\alpha}_{{\bf \theta}_{\ell}}\rangle.
\label{real2}
\end{eqnarray}
From this relation, once we obtain the real-space wave function $|\psi^{\alpha'}_{{\bf \theta}}\rangle$ {\it without twisted boundary phases}, 
all wave functions with different twisted boundary phases can be constructed. 
Therefore, we can directly obtain the elements of the coupling matrix $A(\ell,\ell+1)$, $\langle \psi^{\alpha}_{\theta_\ell}|\psi^{\alpha '}_{\theta_{\ell+1}}\rangle$, from the single-particle wave function without twisted boundary phases. We introduce the real-space-based single-particle wave function as $|\psi^{\alpha}_{{\bf \theta}_\ell}\rangle=\sum_{{\bf r}}c^{\alpha,{\bf \theta}_{\ell}}_{{\bf r},\beta}|{\bf r},\beta\rangle$, where $\beta$ is a sublattice label at ${\bf r}$, and $c^{\alpha,{\bf \theta}}_{{\bf r},\beta}$ is an expansion coefficient. 
By using Eq.~(\ref{real2}), the elements of the matrix in Eq.~(\ref{Cmatrix2}) are then given as 
\begin{eqnarray}
\langle \psi^{\alpha}_{\theta_\ell}|\psi^{\alpha '}_{\theta_{\ell+1}}\rangle=\sum_{{\bf r},\beta}(c^{\alpha,{\bf 0}}_{{\bf r},\beta})^{*}e^{i({{\bar {\bf\theta}}}_{\ell}-{{\bar {\bf \theta}}}_{\ell+1})\cdot {\bf r}}c^{\alpha ', {\bf 0}}_{{\bf r},\beta}.
\end{eqnarray}
Thus, we can obtain all the elements of the coupling matrix $A(\ell,\ell+1)$ from $|\psi^{\alpha}_{\bf 0}\rangle$. 
We can easily obtain $c^{\alpha, {\bf 0}}_{{\bf r},\beta}$ numerically by diagonalizing the real-space Hamiltonian.
Thus, we can calculate each element of the coupling matrix $A(\ell,\ell+1)$, and 
the Chern number $C_{N}$ is obtained using Eq.~(\ref{CNphase3}). 

In the coupling matrix method, as long as we obtain the real-space single-particle wave function even in a system without translational invariance,
 we can obtain $C_{N}$ using Eq.~(\ref{CNphase3}).

\section{Disorder phase diagram of the HHH model}\label{SecIV}
\begin{figure}[t]
\centering
\includegraphics[width=6cm]{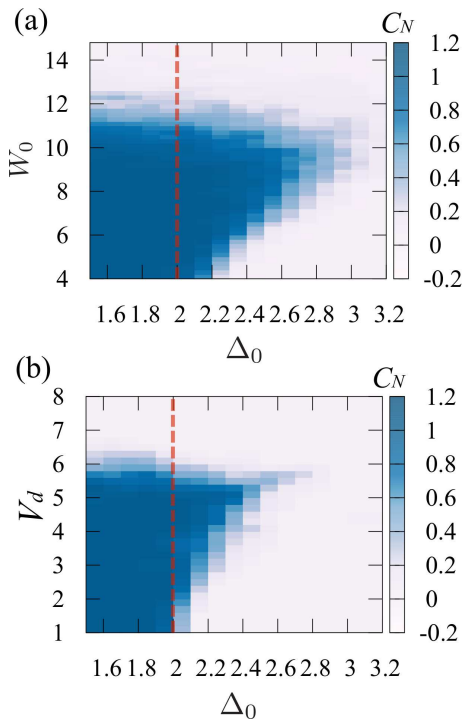}
\caption{(a) On-site random disorder case. We took the average of 40 random on-site samples. 
(b) Quasi-periodic potential case. We took the average of 40 samples with different values of $\phi$.  
The red lines represent the conventional topological phase transition line in the clean limit, $\Delta_{0}=2$.}
\label{Fig2}
\end{figure}
In this section, we calculate the global topological phase diagrams for the two disorder potentials, $V^{\rm I}_{d}$ and $V^{\rm II}_{d}$. 
We apply a periodic boundary condition to the system. In particular, we focus on how the disorder potentials modify the phase boundary between the trivial ($C_{N}=0$) 
and nontrivial topological CI phase ($C_{N}=1$) in the HHH model.
These phases are identified by the Chern number $C_{N}$. 
Qualitatively, for weak disorder the CI phase expects to be stable because a large bulk gap 
protects the CI phase. Large enough disorder works to close the bulk band gap, and leads the disappearance of the CI phase. 
On the other hand, when disorder has a moderate strength where whether the bulk band gap is closed is subtle, 
non-trivial phenomena may appear in the behavior of the bulk band gap.

Using the coupling matrix method introduced in the previous section, we determined the ground-state phase diagram. 
Figure~\ref{Fig2}(a) shows the result for the $V^{\rm I}_{d}$ disorder case. The system size is $(L_{x},L_{y})=(20, 20)$. 
This result indicates that the CI phase is robust up to $W_{0}\sim 11$ for small $\Delta_{0}$, and 
we find that for moderately strong disorder, the regime of the CI phase characterized by $C_{N}\sim1$ is enlarged because of the on-site random disorder; 
the phase boundary is clearly beyond that in the clean limit case $\Delta_{0}=2$, as indicated by the red line in Fig.~\ref{Fig2}(a). 
This phase diagram shows that when we increase $W_{0}$ from zero at a certain value of $\Delta_{0}$ (which is slightly larger than $\Delta_{0} = 2$), 
the system exhibits an interesting transition flow: trivial phase $\to$ CI phase $\to$ trivial phase. 

Next, we consider the quasi-periodic potential case, $V^{\rm II}_{d}$. 
Here, to suppress the boundary effect 
resulting from the fact that the quasi-periodic potential is discontinuous at the boundary, 
we average over $\phi$, where $\phi$ is divided into 40 parts between  $0$ and $2\pi$. 
The result is shown in Fig.~\ref{Fig2}(b). The unit cell lattice size is $(L_{x},L_{y})=(20, 20)$. 
As in the on-site random disorder case, $V^{\rm I}_{d}$, 
the CI phase is robust against the quasi-periodic disorder up to $V_{d}\sim 6$ at $\Delta_{0} < 2$. 
Interestingly, we find that even with the quasi-periodic disorder, the regime of the CI phase is enlarged beyond $\Delta_{0} > 2$ at moderate values of $V_{d}$.
Therefore, the HHH model exhibits the disorder-induced CI phase, which is defined in the $\Delta_{0}>2$ regime, 
for both on-site random disorder and quasi-periodic potential.

\begin{figure}[t]
\centering
\includegraphics[width=7cm]{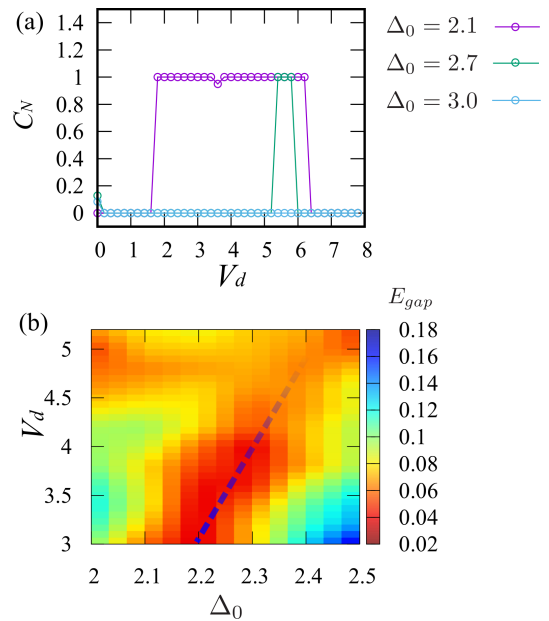}
\caption{(a) Disorder dependence of $C_{N}$ for $\Delta_{0}=2.1\;, 2.7$, and $3$. 
(b) Bulk band gap distribution near the phase boundary between the disorder-induced CI phase and trivial phase.
The blue dashed line is an approximate phase boundary between the CI and the trivial phase at which the bulk band gap is very small.}
\centering
\label{Fig3}
\end{figure}

Furthermore, figure~\ref{Fig3}(a) shows the behavior of $C_{N}$ in detail for the quasi-periodic disorder and typical fixed $\Delta_{0}$ values. 
We capture a clear plateau with $C_{N}=1$ for $\Delta_{0}=2.1$ for moderately strong disorder. 
The result exhibits the clear transition flow: trivial phase $\to$ CI phase $\to$ trivial phase.
This is a disorder-induced CI phase caused by the quasi-periodic disorder, i.e., the revival of the CI phase. 
For larger $\Delta_{0}$, the plateau decreases, as shown in Fig.~\ref{Fig3}(a).
A similar transition flow has been reported in other topological models but not the HHH model \cite{Liu,Meier,Castro,Mondragon,Sriluckshmy,Yamakage}.

In addition, we calculate the bulk band gap near the phase boundary between the CI phase and the trivial phase 
at moderate values of $V_{d}$ for $V^{\rm II}_{d}$ with $\phi=0$. 
Here, the bulk band gap $E_{gap}$ is defined as the energy difference 
between the $(L_{x}L_{y}/2)$-th and $(L_{x}L_{y}/2+1)$-th eigenenergies in the real-space Hamiltonian of the HHH model.
We plot the distribution of the bulk band gap $E_{gap}$ in Fig.~\ref{Fig3}(b). 
The result indicates that $E_{gap}$ is small but finite in the disorder-induced CI regime. 
The transition from the disorder-induced CI phase to the trivial phase seems to originate 
from the conventional gap-closing behavior in typical topological phase transitions: 
gapful $\to$ gapless critical point $\to$ gapful as the parameter is varied. 
See the blue dashed line Fig.~\ref{Fig3}(b). The gap closing tendency persists up to $V_{d}\sim 5$.
For more detail, we also calculate the DOS in the next section. 
\begin{figure}[t]
\centering
\includegraphics[width=8cm]{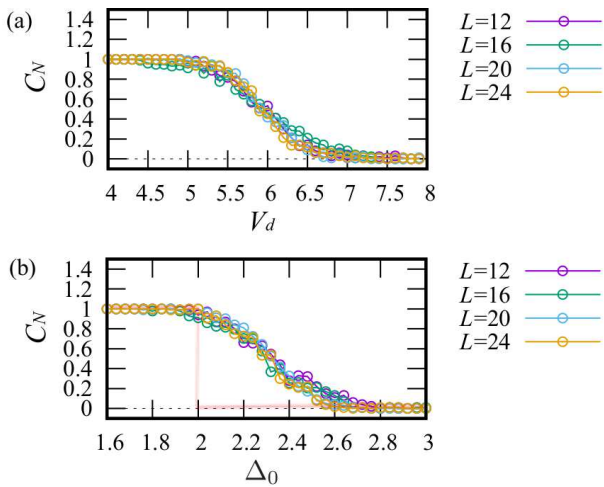}
\caption{System size dependence of $C_{N}$ for (a) $\Delta_{0}=1.6$ 
and (b) $V_{d}=4$. 
The light red line represents the conventional CI phase transition behavior in the clean limit, $\Delta_{0}=2$.
Here, $L_{x}=L_{y}=L$. For all data, we took the average of 40 samples with different values of $\phi$.}
\label{Fig4}
\end{figure}

To consolidate the global phase diagrams in Fig.~\ref{Fig2}, 
we estimate the system size dependence of the coupling matrix method.
We focus on the $V^{\rm II}_{d}$ disorder case. 
Figure~\ref{Fig4}(a) and (b) shows the system size dependence of $C_{N}$ for transitions from the conventional CI phase to the trivial phase 
and from the disorder-induced CI phase to the trivial phase. 
In both cases, the results indicate a weak system size dependence. 
Therefore, our calculation results are expected to be very close to those of the thermodynamic limit.

In addition, we comment that in Fig~\ref{Fig4}(a) and (b), there is a regime where a non-quantized value of $C_{N}$ appears. 
The regime may originates from a consequence of an emergent spatial inhomogeneity of the system. 
Actually, the calculation of the Chern number in a similar situation has been performed using the local Chern marker \cite{Moscot}, 
where the spatial dependence of the local Chern number is shown clearly. 
In appendix, we show for a typical parameter point, sample-to-sample distribution of $C_{N}$ obtained from the coupling matrix method.

\section{Density of states}\label{SecV}
To further confirm the existence of the disorder-induced CI phase, the gap-closing behavior should be studied in detail.
The DOS \cite{Weise} provides clues as to how the bulk band gap closes and also the localization properties.
The DOS is defined as follows:
\begin{eqnarray}
\rho(E_{i})=\frac{1}{N_{D}}\sum_{k}\delta(E_{i}-E_{k}).
\end{eqnarray}
Here, $E_{i}$ and $N_{D}$ are the $i$-th energy eigenvalue 
and the Hilbert space dimension, respectively.
Practically, we calculate $\rho(E_{i})$ 
by setting a small but finite energy window $\Delta E$, in which $E_{i}$ is centered. 

\begin{figure}[t]
\centering
\includegraphics[width=6.5cm]{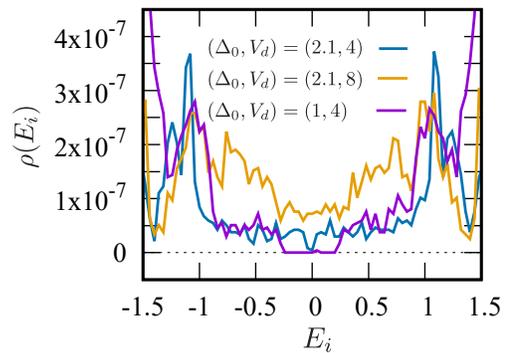}
\caption{DOS. The blue, yellow, and purple lines represent the disorder-induced CI phase $(\Delta_{0},V_{d})=(2.1,4)$, 
trivial phase $(\Delta_{0},V_{d})=(2.1,8)$, and conventional CI phase $(\Delta_{0},V_{d})=(1,4)$, respectively.}
\label{Fig5}
\end{figure}

We consider the quasi-periodic potential $V^{\rm II}_{d}$ of Eq.~(\ref{dis2}) with $\phi=0$ and 
calculate the DOS around the band center ($E_{i}\sim 0$) at typical parameter points in the phase diagram in Fig.~\ref{Fig2}(b). 
Figure~\ref{Fig5} shows the DOS for a system size of $(L_{x},L_{y})=(62,124)$. 
The parameter points are $(\Delta_{0},V_{d})=(2.1,4)$, $(2.1,8)$, and $(1,4)$, 
which correspond to the disorder-induced CI phase, trivial phase, and conventional CI phase, respectively.
For the conventional CI point, $(\Delta_{0},V_{d})=(2.1,4)$, $\rho(E_{i})$ is clearly zero around $E_{i}=0$. 
This means that the conventional CI clearly has a large bulk band gap. 
For the trivial phase point, $(\Delta_{0},V_{d})=(2.1,8)$, $\rho(E_{i})$ is finite even around $E_{i}=0$. 
This indicates that for strong disorder, the trivial phase has no bulk band gap, and the system tends to have metallic properties. 
For the disorder-induced CI point, $(\Delta_{0},V_{d})=(2.1,4)$, $\rho(E_{i})$ is zero in a small region around $E_{i}=0$. 
That is, this result indicates that a small bulk band gap has survived. 
By combining this result and the result of direct bulk band gap calculation in Fig.~\ref{Fig3}(b), 
we can conclude that the disorder keeps the bulk band gap open even though the parameter $\Delta_{0}$ exceeds the value of the CI phase transition point in the clean limit. 
We expect that this mechanism for the emergence of the disorder-induced CI is somewhat different from that of the TAI \cite{Guo2,Groth,YYZhang}.
The TAI is believed to arise from a mobility gap \cite{Shen,YYZhang} at which the bulk band gap vanishes; 
thus, $\rho(E_{i})$ is finite even around $E_{i}=0$ \cite{YYZhang}. 

In addition, before closing this section, we should note that the data at $(\Delta_{0},V_{d})=(2.1,4)$ exhibit sharp peaks at $E_{i}\sim \pm 1$. This indicates the presence of a mobility edge. 
Similar behavior has been reported in another model \cite{Castro}. 
Therefore, in the disorder-induced CI phase, the system spectrum also includes weak localization features. 
This result is reminiscent of the result in Ref. \cite{Prodan}.

\section{Possibility of disorder-induced CI in a one-dimensional reduced model}\label{SecVI}
In the previous sections, we focused on the non-interacting HHH model and found the disorder-induced CI phase.
In this section, we study whether such a phase also exists in an interacting topological system. 
It is generally difficult to numerically simulate the 2D HHH model including interactions. 
Thus, we consider a one-dimensional (1D) model obtained by considering a dimensional reduction of the HHH model.
The 1D model can be derived using the dimensional reduction technique \cite{Kraus}.
A 2D model $H_{\rm 2D}$ is reduced to a 1D model $H_{\rm 1D}(\rho)$ with an adiabatic parameter $\rho$:
\begin{eqnarray}
H_{\rm 2D}=\int\frac{d\rho}{2\pi}H_{\rm 1D}(\rho).
\label{1Dreduce}
\end{eqnarray}
We apply the above formula to the HHH model in Eq.~(\ref{HHH}). We then assume a bosonic system with an on-site interaction and add an on-site random disorder.
The resulting 1D model is given by
\begin{eqnarray}
H_{\rm 1D}(\rho)&=&\sum_{m}\biggl[ -J_{m}(\rho)(b^{\dagger}_{m+1}b_{m}+\mbox{h.c.})\nonumber\\
&+&(\Delta_{m}(\rho)+ W_{m})b^{\dagger}_{m}b_{m} 
+\frac{U}{2}b^{\dagger}_{m}b_{m}(b^{\dagger}_{m}b_{m}-1)\biggr]. \nonumber\\
\label{BRM}
\end{eqnarray}
\begin{figure}[t]
\centering
\includegraphics[width=8cm]{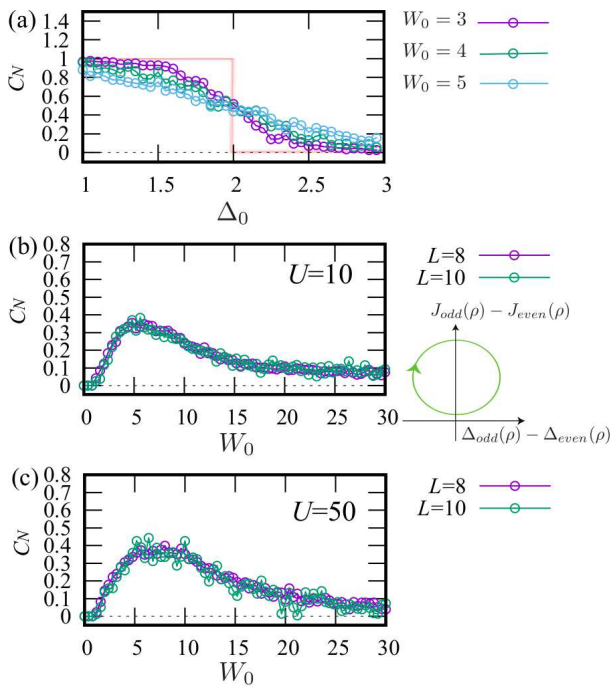}
\caption{(a) Disorder dependence of $C_{N}$ for the non-interacting HHH model with an $x$-site-dependent on-site random disorder. 
The light red line represents the conventional CI phase transition behavior in the clean limit, $\Delta_{0}=2$.
(b) and (c): Disorder dependence of $C_{N}$ for the 1D interacting RM model with $U=10$ and $U=50$. 
In both cases, the particle density is $1/2$, and 
we took the average of $1000$ and $400$ disorder samples for $L=8$ and $L=10$.}
\label{BRM_CN}
\end{figure}
This is a bosonic interacting RM model, where $m$ is a 1D lattice site, $b^{\dagger}_{m}$ ($b_{m}$) is 
a bosonic creation (annihilation) operator at site $m$, $U$ is the on-site interaction, $W_{i}$ is a uniformly distributed random potential 
with a strength $W_{0}$ that is defined as $W_{m}\in [-W_{0}/2,W_{0}/2]$.
Further, the parameters $J_{m}(\rho)$ and $\Delta_{m}(\rho)$ are given by
\begin{eqnarray}
&&J_{m}(\rho) = t_a + 2t_{c} (-1)^{m}\cos(\rho),\\
&&\Delta_{m}(\rho) = [2t_{b}\sin (\rho)-\Delta_{0}](-1)^{m+1}. \label{deltarho}
\label{modelprameter}
\end{eqnarray}
The $y$-direction hopping and on-site staggered potential in the HHH model are transformed into a modulated on-site potential in the 1D model, 
and the NNN hopping is transformed into a modulated part of the hopping in the 1D model. 
The bosonic RM model has already been realized in cold atom experiments \cite{Lohse,Nakajima}. 
The $\Delta_{0}$ term in Eq.~(\ref{deltarho}) can be implemented as an additional staggered lattice potential in real experimental systems.
In real experiments, the Chern number $C_{N}$ in the RM model can be extracted as the shift of the Wannier center in the optical lattice. 

To detect the disorder-induced CI phase as shown in the previous sections, 
we consider a half-filled system (boson particle density $1/2$) and periodic boundary condition, and we employ exact diagonalization \cite{ED0,ED1,ED2}.
To identify the topological phase, we use the Chern number \cite{Fukui} 
parameterized by $\rho$ and the twisted boundary phase $\alpha$ \cite{SLZhu, Kuno2}.

First, for comparison, we calculate the fermionic HHH model in Eq.~(\ref{HHH}) with only an $x$-site-dependent on-site random disorder, 
which is written as $\sum_{(m,n)}W_{m} c^{\dagger}_{(m,n)}c_{(m,n)}$. 
By applying dimensional reduction to Eq.~(\ref{1Dreduce}), 
the model can also be mapped onto the non-interacting RM model with random on-site disorder in one spatial dimension. 
Figure~\ref{BRM_CN}(a) shows the $\Delta_{0}$ dependence of the Chern number for typical moderately strong disorder. 
Here, the topological phase transition becomes somewhat crossover-like in a region centered on the clean limit transition point $\Delta_{0}=2$.
Even at $\Delta_{0}>2$, the averaged value of $C_{N}$ obtained from different disorder samples is finite to some extent; 
this indicates that the CI phase characterized by $C_{N}=1$ is often generated depending on the disorder sample. 
Let us turn to the exact diagonalization results for the interacting bosonic RM model, which are shown in Fig.~\ref{BRM_CN}(b) and (c). 
Here, we set $\Delta_{0}=2.1$ and vary the disorder strength $W_{0}$. 
For interaction strengths of $U=10$ and $50$, 
the average values of $C_{N}$ are fairly large ($\sim 0.4$) for moderately strong disorder,
and the behavior seems independent of the system size and interaction strength $U$.
We expect that the increase tendency of the value of $C_{N}$ in a moderately strong disorder regime indicates the presence of the disorder-induced CI state. 
As in the non-interacting fermion case in Fig.~\ref{BRM_CN}(a), 
the interacting bosonic RM model also exhibits revival of the CI state.
In addition, as mentioned at the end of Sec.~\ref{SecIV}, in appendix, 
we show, for a typical case in Fig.~\ref{BRM_CN}(b), sample-to-sample distribution of $C_{N}$, exhibiting a non-quatized value of $C_{N}$ on average.

\section{Discussion and conclusion}\label{SecVII}
In this study, we investigated the effect of disorder on the CI phase in the HHH model.
Two types of disorder were considered: those represented by the on-site random potential and Aubry--Andre quasi-periodic potential. 
Global phase diagrams were calculated using the coupling matrix method. 
We found that in the HHH model, a disorder-induced CI phase appeared for both types of disorder.
We calculated the bulk band gap and DOS. These calculation revealed that the disorder-induced CI has a small but finite bulk gap. 
This situation is somewhat different from that of the TAI, which originates in a mobility gap. 
Furthermore, we detected the disorder-induced CI phase in an interacting bosonic RM model associated with the HHH model 
by using dimensional reduction.
There, we obtained results similar to those for the non-interacting fermionic HHH model; that is, even the interacting bosonic RM model exhibited
 revival of the CI state induced by random potential disorder at the level of the exact diagonalization calculation. 
This revival event may be tested in real experimental cold atom systems \cite{Lohse, Nakajima} 
because bosonic atom experiments reach a very low-temperature regime and thus can highly suppress finite-temperature effects.
Further, a random potential can be implemented by using a speckle laser \cite{speckle1,speckle2}.
In addition, for the interacting bosonic RM model, a larger system size may be numerically accessible if greater computational resources are used. 
This would be an interesting problem for future work.

All the behavior found in this study may appear in the topological Haldane model, which has been realized in cold atom experiments \cite{Jotzu,Asteria}.
It may be interesting to see how the disorder-induced CI phase appears in this system.
Further, it may be possible to treat the effects of disorder using the replica theory \cite{Edwards},
in which the on-site random disorder potential can be transformed into an effective interaction term composed of different replica fields.
This interaction term may renormalize the mass term (which corresponds to the $\Delta_{0}$ term in the HHH model). 
As a result, it may lead to a phase boundary shift of the topological phase transition; i.e., it may generate the disorder-induced CI phase.
\bigskip

\acknowledgments
Y. K. thanks Y. Takahashi and S. Nakajima for helpful discussions.
Y. K. acknowledges the support of a Grant-in-Aid for JSPS
Fellows (No. 17J00486). 

\appendix
\renewcommand{\theequation}{A.\arabic{equation}}

\section*{Appendix: Sample-to-sample distribution of the Chern number}
\begin{figure}[h]
\centering
\includegraphics[width=6cm]{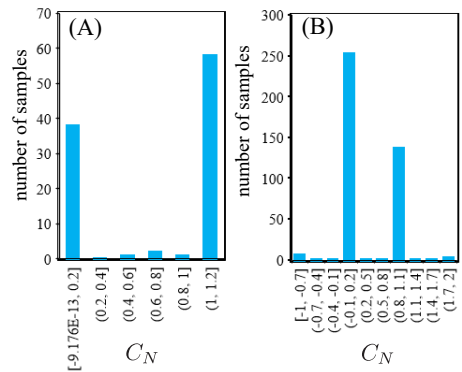}
\caption{(A) Sample-to-sample distribution of $C_{N}$ for $V_d=5.75$ and $L=20$ in Fig.~\ref{Fig4} (a).   
(B) Sample-to-sample distribution of $C_{N}$ for $W_0=5$, $U=10$ and $L=10$ in Fig.~\ref{BRM_CN}(b).}
\label{CN_dis}
\end{figure}

In this appendix, we show a typical sample-sample distribution of $C_{N}$ 
when the value of $C_{N}$ takes a finite non-quantized value under moderately strong disorder.
Figure~\ref{CN_dis} (A) indicates the sample-to-sample distribution of $C_{N}$ for $V_d=5.75$ and $L=20$ in Fig.~\ref{Fig4} (a).
Here, we took 100 disorder samples. Each values of $C_{N}$ almost take 0 or 1, not take fractional value. 
The $C_{N}=1$ case (CI phase) occurs more than the $C_{N}=0$ case. The average value is $0.609$.
In addition, figure~\ref{CN_dis} (B) indicates the sample-to-sample distribution of $C_{N}$ for $W_{0}=5.75$ and $L=10$ in Fig.~\ref{BRM_CN} (b).
Here, we took 400 disorder samples. Same as the result in Fig.~\ref{CN_dis} (A), each values of $C_{N}$ almost take 0 or 1 though there are some other values. 
The $C_{N}=0$ case (trivial phase) occurs more than the $C_{N}=1$ case. The average value is $0.401$. 
Accordingly, both results in Fig.~\ref{CN_dis} indicate that when the value of $C_{N}$ takes a finite non-quantized value under moderately strong disorder, 
the system tends to exhibit the CI phase or the trivial phase depending on the sample.

 
\end{document}